\documentclass[prl,twocolumn,showpacs]{revtex4}
\usepackage[dvips]{graphicx}

\begin{document}
 
\title{Instability of isolated triplet excitations on the 
Shastry-Sutherland lattice (SSL)}
\author{A. Fledderjohann, K.-H. M{\"u}tter}
\affiliation{Physics Department, University of Wuppertal, 42097 Wuppertal, Germany}
%
\begin{abstract}
Configurations of singlets and triplets on the SSL have been
proposed in the literature as variational ground states of
the Shastry-Sutherland model at fixed magnetization $M$. We
prove, that isolated triplet excitations on the SSL are 
unstable if the coupling $\alpha$ falls below a critical value 
$\alpha_c\approx 2.0$. The instability should be visible in
the compound $\mathrm{SrCu_2(BO_3)_2}$ where a coupling $\alpha^*=1.48$
is realized.
\end{abstract}
\pacs{75.10.-b, 75.10.Jm}
\maketitle                                                                         
 
\section{Introduction}

\begin{figure}[ht!]
\centerline{\includegraphics[width=7.0cm,angle=0]{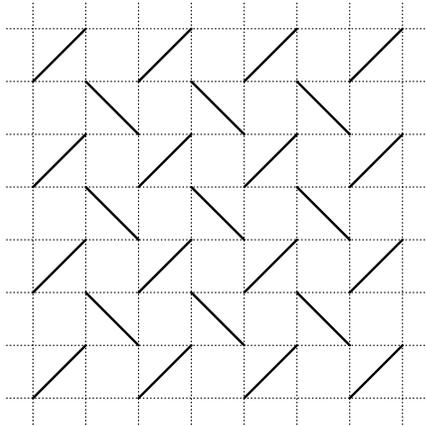}}
\caption{The couplings in the Shastry-Sutherland model. Nearest
and next-nearest neighbor couplings are represented by dotted
and solid lines, respectively. The diagonal bonds define the 
Shastry-Sutherland lattice (SSL).}
\label{fig1}
\end{figure}

The discovery of plateaus in the magnetization curve $M=M(B)$ of the
compound $\mathrm{SrCu_2(BO_3)_2}$  \cite{kageyama99,onizuka00,nojiri99}
at rational values of the magnetization
$M/M_S=1/3,1/4,1/8$ ($M_S=1/2$) has led to intensive investigations
of the ground state properties of the Shastry-Sutherland model 
\cite{shastry81} with Hamiltonian
\begin{eqnarray}
\label{h0}
H & = & \sum_{\langle {\bf x},{\bf y}\rangle}{\bf S}({\bf x})
{\bf S}({\bf y})+\alpha\sum_{\langle\langle {\bf x},{\bf y}\rangle
\rangle}{\bf S}({\bf x}){\bf S}({\bf y})\,.
\end{eqnarray}
The spin-1/2 couplings extend over nearest ($\langle {\bf x},{\bf y}
\rangle$)
and specific next-nearest ($\langle\langle {\bf x},{\bf y}\rangle\rangle$)
neighbor couplings, which define the SSL,
shown in Fig.~\ref{fig1}.

In the absence of an external field ($M=0$) the ground state can be
represented by a product of singlet states on the SSL if $\alpha$
exceeds a critical value $\alpha_c$, which has been found to be
$\alpha_c=1.43$ in Ref.~\onlinecite{miyahara99,weihong01}. 
This value is very close to
the coupling $\alpha^*=1.48$, which seems to be realized in the
compound $\mathrm{SrCu_2(BO_3)_2}$.
In order to describe the magnetic properties of the compound -- in
particular for fixed magnetizations $M=1/4,1/6,1/8,1/12,1/16$ where
plateaus are expected or found -- several ordering patterns have
been proposed and discussed in the literature.
\cite{kageyama99,onizuka00,nojiri99,miyahara99,momoi00,fukumoto00,miyahara00}
They all have in
common that in the sector with total spin $S$, $N_T=S$ triplets
and $N_S=N/2-N_T$ singlets are distributed over the SSL with $N$ 
sites. Typical examples of these singlet-triplet configurations
are shown in Ref \onlinecite{miyahara00}. 
In contrast to the pure singlet configuration
(Fig.~\ref{fig1}) the singlet-triplet configurations are not eigenstates
of the Hamiltonian (\ref{h0}). Moreover, it has been shown in our
previous paper \cite{fle02} that the formation of triplets 
on the SSL is
disadvantegeous if the coupling $\alpha$ falls below a certain
critical value $\alpha_c(M)$, depending on $M$.
In particular, the singlet-triplet
configurations proposed in Fig.8 of Ref.~\onlinecite{miyahara00} 
cannot be
considered as adequate variational states for $M=1/8$, since
$\alpha_c(M=1/8)=2.3>\alpha^*$. As an alternative, we proposed in
Ref.~\onlinecite{fle02} (Fig.4) a ``monomer-dimer configuration'' 
built up
from monomers (i.e. isolated spin-up states) and dimers (i.e.
singlets) on nearest and next-nearest neighbor bonds.


It is the purpose of this Letter, to prove that the instability
of isolated triplet excitations on the SSL is a quite general
feature for $\alpha<2$ and low magnetizations $M\leq 1/8$.


 
\section{Singlet-triplet versus monomer-dimer configurations}
 
\begin{figure}[ht!]
\centerline{\includegraphics[width=6.5cm,angle=0]{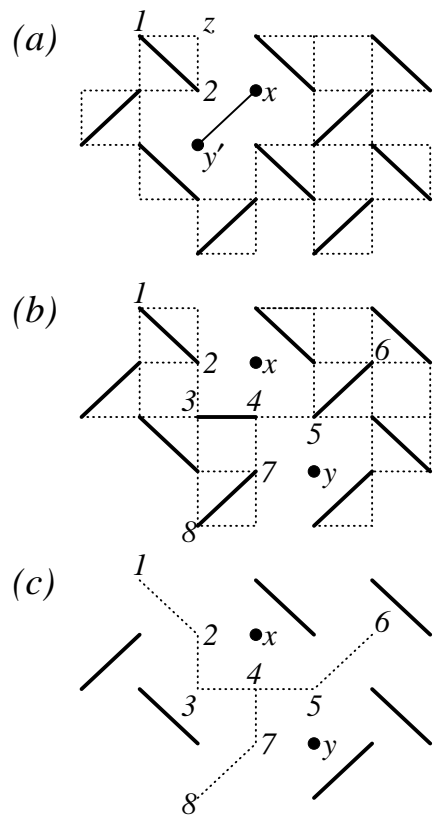}}
\caption{(a) An isolated triplet on the SSL. The singlets
(solid lines) and the dotted
couplings define the antiferromagnetic cluster $(K_a,\nu_a)$.
(b) Break up of the triplet into isolated monomers at sites
{\bf x} and {\bf y} and the antiferromagnetic cluster $(K_b,\nu_b)$.
(c)  The antiferromagnetic 8-site cluster, which leads to the
energy difference (\ref{delta_8}).}
\label{fig2}
\end{figure}

The variational states $|K,\nu\rangle$ which are constructed from
the monomer-dimer configurations generalize the familiar
singlet-triplet configurations on the SSL
in the following sense: In the sector with total spin $S$
we have $\nu=2S$ monomers -- i.e. isolated spin-up states
$|{\bf x}_j+\rangle$ $j=1,\ldots,\nu$ at sites ${\bf x}_1\ldots
{\bf x}_{\nu}$. They form the ferromagnetic cluster ($\nu$).
All remaining sites ${\bf x},{\bf y},\ldots$ are occupied with
$(N-\nu)/2$ singlets (dimers) $[{\bf x},{\bf y}]$.
They form the antiferromagnetic cluster ($K$).
Each monomer and dimer configuration defines a variational
state:
\begin{eqnarray}
\label{var2_ansatz}
|K,\nu\rangle & = & \prod_{j=1}^{\nu}|{\bf x}_j+\rangle
\prod_{\langle {\bf x},{\bf y}\rangle\in K}[{\bf x},{\bf y}]\,.
\end{eqnarray}
The expectation value of Hamiltonian (\ref{h0}) between
these states turns out to be \cite{fle02}
\begin{eqnarray}
\label{ansatz0}
\langle K,\nu|H|K,\nu\rangle & = & -\frac{3}{4}\left( N_1^{(0)}(K)
+\alpha N_2^{(0)}(K)\right)\nonumber\\
 & & +\frac{1}{4}\left( N_1^{(1)}(\nu)
+\alpha N_2^{(1)}(\nu)\right)
\end{eqnarray}
where $N_1^{(0)}(K)$ and $N_1^{(1)}(\nu)$ are the numbers of
nearest neighbor singlets on ($K$) and monomer pairs on ($\nu$),
respectively. $N_2^{(0)}(K)$ and $N_2^{(1)}(\nu)$ denote the
corresponding numbers on the next-nearest neighbors according
to Fig.~\ref{fig1}. The singlet-triplet configurations on the
SSL are special monomer-dimer configurations
with:
\begin{eqnarray}
N_1^{(0)}=0,\,\,N_2^{(0)}=\frac{N-\nu}{2} & , &
N_1^{(1)}=0,\,\,N_2^{(1)}=\frac{\nu}{2}\,,
\end{eqnarray}
i.e. we have a maximum number of dimers (singlets) and monomer 
pairs (triplets) on the SSL.
Due to the geometry of the SSL it is not
possible to maximize the number of singlets $N_2^{(0)}$ and to
minimize the number of triplets $N_2^{(1)}$ at the same time.
Here begins the problem with triplet excitations on the SSL: 
For smaller values
of $\alpha$ [$\alpha<\alpha_c(M)$], the break up of the
triplets into well separated monomer pairs is favored energetically.
For $M\leq 1/8$ -- where the triplets on the SSL
are well isolated (cf. Figs. 8,9,10 in  
Ref.~\onlinecite{miyahara00}) --
we can study the break up of each triplet and the change of the
local environment in the monomer-dimer configuration directly,
as is shown in Fig.~\ref{fig2}(a),(b). The configuration ($K_a,\nu_a$) in
Fig.~\ref{fig2}(a) shows a triplet at sites ${\bf x}$ and
${\bf y}\prime$ embedded in a sea of 
singlets on the SSL. They are represented by the solid lines
in Fig.~\ref{fig2}(a).
The triplet is broken up in a pair of monomers at sites ${\bf x}$
and ${\bf y}$ separated by a 
``knight's move'' [cf. Fig.~\ref{fig2}(b)]. Comparing the energies of the
configurations ($K_a,\nu_a$) and ($K_b,\nu_b$) we have to 
decrease $N_2^{(0)}$ and $N_2^{(1)}$ and to increase $N_1^{(0)}$
by one unit:
\begin{eqnarray}
\label{delta_E}
\Delta E & = & \langle K_b,\nu_b|H|K_b,\nu_b\rangle -
\langle K_a,\nu_a|H|K_a,\nu_a\rangle\nonumber\\
 & = & \frac{1}{2}
\left(\alpha-\frac{3}{2}\right)\,.
\end{eqnarray}
The energy difference (\ref{delta_E}) is negative for $\alpha<3/2$.
Therefore, the break up of the triplet is favored in this
regime. We will see in the next section, that this phenomenon
occurs already at larger values of $\alpha$, if we improve
our variational ansatz.

 
\section{Magnetic order and the frozen monomer approximation}

The monomer-dimer configurations are not eigenstates of the
Hamiltonian (\ref{h0}). However, it might happen that the
ground state is governed by a specific distribution ($\nu$)
of ``frozen'' monomers. They define the magnetic ordering
of the ground state. 
The variational ansatz (\ref{var2_ansatz}) will be improved
if we substitute the product of dimers
\begin{eqnarray}
\label{var_ansatz}
\prod_{\langle{\bf x},{\bf y}\rangle\in K}
[{\bf x},{\bf y}] & \rightarrow & \Psi(K)
\end{eqnarray}
by the ground state $\Psi(K)$ of the antiferromagnetic cluster 
Hamiltonian $H(K)$:
\begin{eqnarray}
H(K)\Psi(K) & = & E(K)\Psi(K)\,.
\end{eqnarray}
$H(K)$ contains all nearest and
next-nearest neighbor couplings on the antiferromagnetic
cluster $K$. In Fig.~\ref{fig2}(a),(b) they are represented by the 
solid singlet lines and the dotted lines connecting them.

The variational ansatz (\ref{var_ansatz}) yields an upper
bound on the ground state energy $E_0(M=\nu/2N)\leq E(K,\nu)$ in the
sector with magnetization $M$, where
\begin{eqnarray}
E(K,\nu) & = & \frac{1}{4}\left(N_1^{(1)}(\nu)+\alpha N_2^{(1)}
(\nu)\right)+E(K)\,.
\end{eqnarray}
Concerning the triplet-singlet configuration in Fig.~\ref{fig2}(a),
the ground state $\Psi(K_a)$ of the antiferromagnetic cluster
Hamiltonian $H(K_a)$ is again given by the product of singlets
on the SSL.
Note that each singlet (e.g. $[1,2]$) is accompanied
by two nearest neighbor couplings ${\bf S}(1){\bf S}(z)$,
${\bf S}(2){\bf S}(z)$ such that the total spin operator
${\bf S}(1)+{\bf S}(2)$ acts on the singlet $[1,2]$. For this reason
the ground state energy cannot be lowered in the case ($K_a,\nu_a$).

\begin{figure}[ht!]
\centerline{\includegraphics[width=8.0cm,angle=0]{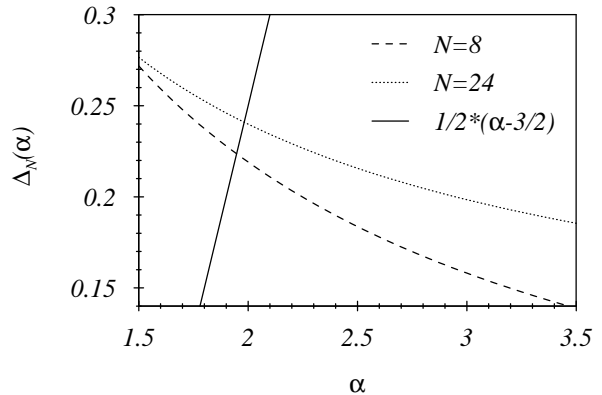}}
\caption{The energy differences (\ref{delta_E}),(\ref{delta_8})
and (\ref{delta_24}). The intersection points define the
critical coupling $\alpha_c$ where the break up of the triplet
sets in.}
\label{fig3}
\end{figure}

The situation is different in configurations with isolated
monomers. The ground state $\Psi(K_b)$ of the antiferromagnetic
cluster is not identical with the dimer product shown in
Fig.[2(b)]. If we apply here the couplings ${\bf S}(2){\bf S}(3)$,
${\bf S}(4){\bf S}(5)$ and ${\bf S}(4){\bf S}(7)$ onto the
nearest neighbor singlet $[3,4]$ a new state is created.
Therefore, the ground state energy $E(K_b)$ of the antiferromagnetic
cluster $K_b$ will be lower than the expectation value of the
corresponding monomer-dimer configuration in Fig.~\ref{fig2}(b). This
will shift the critical value $\alpha_c$ where the triplet
excitation becomes unstable to larger values $\alpha_c>3/2$.
Indeed, we can derive an improved lower bound for $\alpha_c$,
if we substitute the dimers $[1,2][3,4][5,6][7,8]$ by an
antiferromagnetic cluster with 8 sites $1,\ldots,8$ as shown
in Fig.~\ref{fig2}(c).
The ground state energy $E_8(\alpha)$ of the 8-site cluster
is lower than the energy of the corresponding dimer configuration
in Fig.~\ref{fig2}(b) by an amount
\begin{eqnarray}
\label{delta_8}
\Delta_8(\alpha) & = & -\frac{3}{4}(3\alpha+1)-E_8(\alpha)\,;
\end{eqnarray}
$\Delta_8(\alpha)$ is shown in Fig.~\ref{fig3}. The straight line
$\Delta E=(\alpha-3/2)/2$ represents the energy difference
(\ref{delta_E}).
Therefore, the intersection point
\begin{eqnarray}
\label{critical}
\alpha_c & = & 1.95
\end{eqnarray}
yields the critical value, where the instability of the triplet
sets in.
Enlarging the antiferromagnetic surrounding of the frozen
monomer pair in Fig.~\ref{fig2}(c) leads to an increase of
the critical value (\ref{critical}). However, the effect is
small: a calculation with a 16-site cluster -- containing
the 8-site cluster -- yields $\alpha_c=1.97$.

The changes in the magnetic order from an isolated triplet on the
SSL to a pair of monomers separated by a ``knight's move'' can be 
tested by measuring the spin-spin structure factors, which are
obtained from spin-spin correlators $S_3({\bf x})S_3({\bf y})$.

The expectation values of the spin-spin correlators can be read off
from the monomer-dimer configurations:
\begin{eqnarray}
\label{ssc}
\langle K\nu|S_3({\bf x})S_3({\bf y})|K\nu\rangle & = &
\frac{1}{4}\left\{\delta^{(1)}({\bf x},{\bf y})-
\delta^{(0)}({\bf x},{\bf y})\right\}\nonumber\\
 & &
\end{eqnarray}
where $\delta^{(1)}({\bf x},{\bf y})=1$ and 
$\delta^{(0)}({\bf x},{\bf y})=1$ only if $({\bf x},{\bf y})$
coincides with nearest or next-nearest neighbor bonds on $(K,\nu)$
occupied with a monomer pair or a dimer, respectively. In all
other cases $\delta^{(j)}({\bf x},{\bf y})=0$.
Note in particular, that all correlators with one site ${\bf x}$
on the ferromagnetic cluster $\nu$ and the other site ${\bf y}$
on the antiferromagnetic cluster $K$ vanish.

In the frozen monomer approximation with an antiferromagnetic
cluster wavefunction $\Psi(K)$, (\ref{ssc}) can be extended to
\begin{eqnarray}
\label{ssc2}
\hspace{-2.5cm}
\langle \Psi(K),\nu|S_3({\bf x})S_3({\bf y})|\Psi(K),\nu\rangle & = &
\end{eqnarray}
\vspace{-0.8cm}
\begin{eqnarray}
 & & \frac{1}{4}\delta^{(1)}({\bf x},{\bf y})+
\langle\Psi(K)|S_3({\bf x})S_3({\bf y})|\Psi(K)\rangle
\delta\left({\bf x},{\bf y}\in K\right)\,.\nonumber
\end{eqnarray}
Note that (\ref{ssc}) and (\ref{ssc2}) differ in the correlations
$S_3({\bf x})S_3({\bf y})$, ${\bf x},{\bf y}\in K$ on the
antiferromagnetic cluster $K$.


\section{Coexistence of stable and unstable triplet excitations on 
the SSL for $M=1/6$}

\begin{figure}[ht!]
\centerline{\includegraphics[width=7.0cm,angle=0]{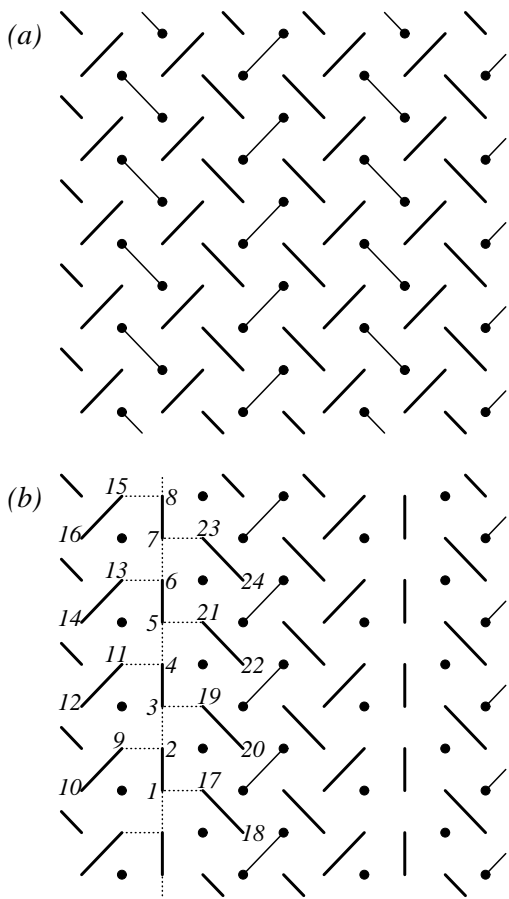}}
\caption{(a) Singlet-triplet configuration ($K_a,\nu_a$) on the SSL for 
$M=1/6$.
(b) Break up of triplets along a stripe with an antiferromagnetic
24-site cluster ($K_b,\nu_b$).}
\label{fig4}
\end{figure}

The instability discussed so far only concerns ``isolated'' triplets
on the SSL. The density of triplets $N_T/N=M$ is controlled by the
magnetization $M$ and raises the question,  
what happens in the cases
$M=1/4$ and $M=1/6$. A typical singlet-triplet configuration 
for $M=1/4$ on the
SSL is given in Fig.~6(a) of Ref.~\onlinecite{miyahara00}. Breaking 
up a triplet in these
configurations produces additional monomer pairs, which increases
the energy, i.e. a high density of triplets prevents their instability.
In the case $M=1/6$, the triplets on the SSL form stripes 
as is shown in Fig.~\ref{fig4}(a).
Interesting enough, each second stripe of the triplets
can break up as is demonstrated in Fig.~\ref{fig4}(b). Comparing  the energies
of the configurations $(K_a,\nu_a)$ and $(K_b,\nu_b)$, we find again, that
the energy difference (\ref{delta_E}) per triplet changes sign for 
$\alpha_c(M=1/6)=3/2$.

Note, that the isolated monomers along one stripe in Fig.~\ref{fig4}(b) are
accompanied with a quasi-onedimensional antiferromagnetic cluster
$K$. We have computed the ground state energy of the 24 site clusters
$E_{24}(\alpha,M=1/4)$, which surrounds the 4 pairs of monomers
arising from the break up of 4 triplets on the SSL,
shown in Fig.~\ref{fig4}(b). The energy difference (per triplet)
\begin{eqnarray}
\label{delta_24}
\Delta_{24}(\alpha,1/6) & = & \frac{1}{4}\left\{-\frac{3}{4}
(8\alpha+4)-
E_{24}(\alpha,1/6)\right\}
\end{eqnarray}
is plotted in Fig.~\ref{fig3}. It meets the straight line
$(\alpha-3/2)/2$ at
\begin{eqnarray}
\alpha_c(M=1/6) & = & 1.985\,.
\end{eqnarray}
This improved lower bound for $\alpha_c(M=1/6)$ differs from
the lower bound we found in Ref.~\onlinecite{fle02}. There we
did not realize, that a mixture of triplets on the SSL and
isolated monomers -- as shown in Fig.~\ref{fig4}(b) -- lowers
the energy more efficiently.


\section{Discussion and conclusions}

In this Letter, we have investigated the stability of triplet
excitations on the Shastry-Sutherland lattice. Our
results can be summarized as follows:

(i) Isolated triplets turned out to be unstable if the ratio
of next-nearest to nearest neighbor couplings is smaller
than $\alpha<\alpha_c\simeq 2.0$. This implies that the
magnetic properties of $\mathrm{SrCu_2(BO_3)_2}$ -- where the coupling
ratio is $\alpha^*\simeq 1.48$ -- cannot be understood in
terms of isolated triplet excitations on the SSL, if the
magnetization $M$ is small enough ($M\leq 1/8$). This 
statement holds in particular for the triplet ordering
patterns for $M=1/8,1/12,1/16$ in Figs. 8,9,10 of
Ref.~\onlinecite{miyahara00}.

(ii) For $M=1/6$ [Fig.~7(a) in Ref.~\onlinecite{miyahara00}]
the triplets form stripes on the SSL. It turns out that the
triplets on each second stripe break up into isolated monomers
if $\alpha\leq\alpha_c(M=1/6)=1.985$.

(iii) The triplets on the SSL in the configuration for $M=1/4$
[Fig.6(a) in Ref.~\onlinecite{miyahara00}] appears to be stable for
$\alpha=\alpha^*$. This demonstrates, that triplet excitations
on the SSL are stabilized if the triplet density is high enough.

Breaking up the isolated triplet into two monomers (spin-up
states) at sites {\bf x} and {\bf y} separated by a ``knight's
move'' [Fig.~\ref{fig2}(c)] lowers the energy of the new
configuration in two ways:
\begin{enumerate}
\item
There is a ``classical'' effect, which results from the
rearrangement of singlets and triplets on $(K_a,\nu_a)$ 
[in Fig.~\ref{fig2}(a)]
into the monomer-dimer configuration $(K_b\nu_b)$ [in 
Fig.~\ref{fig2}(b)].
\item
There is a ``quantum'' effect produced by the antiferromagnetic
cluster, which surrounds the monomers at sites {\bf x} and {\bf y}
[Fig.~\ref{fig2}(c)].
\end{enumerate}


\acknowledgments

We are indebted to M. Karbach for a critical
reading of the manuscript.


 
\end{document}